\documentclass[
reprint,
superscriptaddress,
amsmath,amssymb,
aps,
prab,
floatfix,
]{revtex4-2}

\usepackage{graphicx}%
\usepackage{dcolumn}%
\usepackage{bm}%
\usepackage[english]{babel}
\usepackage[utf8]{inputenc}
\usepackage[mathlines]{lineno}%
\usepackage{siunitx}
\usepackage[dvipsnames]{xcolor}
\usepackage{xspace}
\usepackage{todonotes}
\usepackage{hyperref}%
\hypersetup{
    colorlinks=true,
    linkcolor=blue,
    filecolor=magenta,      
    urlcolor=cyan,
}
\makeatletter
\renewcommand{\fnum@figure}{\normalfont FIG. \thefigure}
\renewcommand*{\@caption@fignum@sep}{\normalfont . }
\makeatother

\newcommand{\rev}[1]{{\color{black}#1}}
\newcommand{\microm}{µm\xspace}

\newcommand{\rms}{\mathrm{rms}}

\newcommand{\IntLaser}{\mathcal{I}}
\newcommand{\divRMS}{\theta_{x}}

\newcommand{\pertpar}{\alpha}

\DeclareSIUnit\mrad{mrad}
\DeclareSIUnit\urad{\si{\micro\radian}}
\DeclareSIUnit\pixel{pixel}

\begin{document}

\preprint{APS/123-QED}

\title{All-optical source size and emittance measurements of laser-accelerated electron beams}%

\author{F.~C.~Salgado}
\email{felipe.salgado@uni-jena.de}
\author{A.~Kozan}
\author{D.~Seipt}
\author{D.~Hollatz}
\author{P.~Hilz}
\author{M.~Kaluza}
\author{A.~S{\"a}vert}
\author{A.~Seidel}
\author{D.~Ullmann}
\author{Y.~Zhao}
\author{M.~Zepf}
\affiliation{Institute of Optics and Quantum Electronics, Friedrich-Schiller-Universit{\"a}t, Max-Wien-Platz 1, 07743 Jena, Germany}
\affiliation{Helmholtz-Institut Jena, Fr{\"o}belstieg 3, 07743 Jena, Germany}
\affiliation{GSI Helmholtzzentrum f{\"u}r Schwerionenforschung, Planckstraße 1, 64291 Darmstadt, Germany}
\date{\today}%
 
\begin{abstract}
Novel schemes for generating ultra-low emittance electron beams have been developed in past years and promise compact particle sources with excellent beam quality suitable for future high-energy physics experiments and free-electron lasers. Recent theoretical work has proposed a laser-based method capable of resolving emittances in the sub 0.1 mm mrad regime, by modulating the electron phase-space ponderomotively. Here we present a first experimental demonstration of this scheme using a laser wakefield accelerator. The observed emittance and source size is consistent with published  values. We also show calculations demonstrating that tight bounds on the upper limit for emittance and source size can be derived from the ‘laser-grating’ method even in the presence of low signal to noise and uncertainty in laser-grating parameters.

\end{abstract}

\maketitle

\section{Introduction}

Laser wakefield accelerators (LWFA) offer an effective alternative for efficiently accelerating electrons to GeV energies within a compact footprint, thanks to their substantial acceleration gradient ranging from 10 to 100~GV/m~\cite{Esarey.2009, Gonsalves.2019}. In contrast, radio-frequency (RF) technology demands several kilometers to achieve GeV-energy electron bunches, relying on electric field gradients of 10 to 50~MV/m. The small size and high quality of the electron beams generated by the laser plasma accelerators make them an attractive driver for free-electron lasers (FEL)~\cite{Wang.2021, Fuchs.2009, Huang.2012, Maier.2012}, electron-positron particle colliders~\cite{Leemans.2009, Schroeder.2010, alegrocollaboration2019advanced, Assmann.2020}, and compact synchrotron-like~\cite{Rousse.2004, Jaroszynski.2006, Kneip.2010} and Thomson sources~\cite{Geddes.2015, Powers.2014}. The quality of such sources, such as their high brightness and coherence, is directly linked to the quality of the electron beam fed into the accelerator. The normalized brightness $B_n$ of a beam is often expressed by the beam current and the transverse beam emittance as~\cite{Cianchi.2016}:
\begin{equation}
    B_n = \frac{2}{\pi}\frac{I}{\pi \, \epsilon_{n_x} \, \epsilon_{n_y}} \,,
    \label{eq. brillance definition}
\end{equation}
where $I$ is the beam current, and $\epsilon_n$ are the transverse normalized emittances in both $x-$ and $y-$ directions.

The emittance of electron beams accelerated by lasers relies on the chosen injection method, such as ionization~\cite{Pak.2010, McGuffey.2010} or downramp injection~\cite{Chien.2005}. These methods determine the phase-space volume into which electrons are injected and the field perturbations affecting them. Consequently, these techniques enable accelerated electron bunches to attain transverse emittances of a similar order of magnitude (e.g., $\epsilon_n < 0.2~\pi$~mm~mrad) as those observed in radio-frequency linear accelerators~\cite{Weingartner.2012}.

To enhance the beam quality of laser-plasma accelerators, recent proposals involve methods utilizing plasma photocathodes and similar techniques~\cite{Hidding.2012, Li.2013, Wittig.2015, Xu.2014, Manahan.2017}. These approaches aim to decouple the plasma wave excitation from electron bunch generation, resulting in a low initial momentum spread and improved confinement of accelerated electrons. Consequently, transverse emittances as low as $\epsilon_n < 0.1$~mm~mrad are predicted~\cite{Hidding.2012}. Achieving such low emittance values is a significant advance in realizing the high brilliance beams required for the next generation of particle colliders~\cite{Leemans.2009, Foster.2023}.

Various methods, including transverse deflecting structures (TDS)~\cite{Akre.2001, Behrens.2014, Dolgashev.2014, Floettmann.2014, Maxson.2017, Tan.2019, Marchetti.2021}, quadrupole/solenoid scans~\cite{Weingartner.2012, Prat.2014, Ji.2019}, and pepper pot (PP) masks~\cite{Fritzler.2004, Shanks.2009, Sears.2010, Brunetti.2010, Manahan.2014}, have been employed for measuring the emittance of electron beams. Nevertheless, accurately measuring the small emittance beams, as predicted by methods like the plasma photocathode scheme, remains challenging due to the requirement for large and complex experimental setups. \rev{For instance, to measure nanometer-scale source sizes, Shintake monitors can be employed~\cite{Tenenbaum.1999, Yan.2012}. These monitors utilize Compton scattering signals generated from the interaction of the electron beam with laser interference fringes to deduce the beam waist. However, this method requires scanning the phase of the interference fringes of the laser relative to the electron beam over multiple shots.}

Recently, a solution to improve the measurement quality of low-emittance beams has been proposed based on an  all-optical method~\cite{Seidel.2021}. This method, namely the laser interference (grating) method, uses a grating-like interference pattern structure created by two focused lasers under an angle $\vartheta$ that modulates the electron beam phase space. As the electron beam propagates through the interference pattern, the ponderomotive force associated with this pattern pushes the electrons away from regions of  high-intensity towards the low-intensity areas of the grating pattern, consequently, modulating the electron beam's momentum distribution~\cite{Seidel.2021}. 
The sensitivity of the method is dependent on the spatial frequency of the interference pattern and the strength of the modulations generated in the phase space of the electron beam, allowing  beam waists spanning from \si{\nm} to \si{\um} sizes to be resolved.

\section{Theory}
\label{theory laser interference method}

\subsection{Beam emittance definitions}
In general, the normalized emittance $\epsilon_n$ is defined as~\cite{Antici.2012}:
\begin{equation}
    \epsilon_n^2 = \, \langle x^{2} \rangle \langle \beta^2 \, \gamma^2 \, x^{\prime \,2} \rangle - \langle x \, \beta \, \gamma \, x^{\prime} \rangle^2 \,,
    \label{eq. normalized emittance definition}
\end{equation}
where $\beta = v/c$ is the normalized velocity of the particle, $\gamma$ is the particle relativistic factor, $x$ represents the particle's transverse position, and $x^\prime$ stands for its divergence. The notation $\langle \cdot \rangle$ represents the second momentum of the quantity within the brackets. In the case of drift without collective effects, a negligible correlation between energy and transverse position is observed. Consequently, eq.~\eqref{eq. normalized emittance definition} can be written as:
\begin{equation}
    \epsilon_n^2 = \, \langle \beta^2 \, \gamma^2 \rangle \langle x^2 \rangle\langle x^{\prime \, 2} \rangle - \langle \, \beta \, \gamma\rangle^2  \, \langle x  x^{\prime} \rangle^2 \,.
    \label{eq. normalized emittance definition without correlation energy-position}
\end{equation}

Now, if one substitutes the definition of the energy spread $\sigma_E/E$,
\begin{equation}
    \left(\frac{\sigma_E}{E} \right)^2 = \, \frac{ \langle \beta^2 \, \gamma^2 \rangle - \langle \beta \, \gamma \rangle^2 }{\langle \gamma \rangle^2} \,,
\end{equation}
into eq.~\eqref{eq. normalized emittance definition without correlation energy-position}, and assumes relativistic electrons, $\beta \approx 1$, the normalized emittance definition becomes:
\begin{equation}
    \epsilon_n^2 = \, \langle \gamma \rangle^2 \left[ \left(\frac{\sigma_E}{E} \right)^2 \, \langle x^2 \rangle \, \langle x^{\prime\,2} \rangle + \epsilon_{\rms}^2\right] \, ,
    \label{eq. norm emittance with energy spread}
\end{equation}
where $\epsilon_{\rms}$ is the root-mean-square (rms) emittance, also known as geometric emittance~\cite{Floettmann.2003}, which is defined as:
\begin{gather}
    \epsilon_{\rms}^2 = \, \langle x^2 \rangle \langle x^{\prime\,2} \rangle - \langle x x^{\prime} \rangle^2 \, \approx\, \sigma_x^2 \, \divRMS^{2} \,,
    \label{eq. rms emittance definition}
\end{gather}
\rev{where $\sigma_x^2$ denotes the rms beam waist, while $\divRMS = \sigma_{x^{\prime}}$ stands for the uncorrelated rms beam divergence in the x-direction (with beam propagation along the z-direction). With this definition, the rms emittance of the beam is equivalent to the beam emittance in the x-direction as $\epsilon_{\rms}=\epsilon_x$. We consider that the beam propagates in z-direction in the entire manuscript.}

\rev{As the electron beam propagates within the plasma channel in the laser wakefield accelerator, it experiences strong focusing forces, maintaining a very small beam waist while exhibiting a large angular spread~\cite{Esarey.2009}. Upon approaching the plasma-vacuum interface, a downramp density transition of a few-hundreds of micrometers occurs, weakens the focusing forces acting on the beam. This results in an increase of the beam waist but a decrease of its divergence, preserving the beam emittance~\cite{Sears.2010, Weingartner.2012}. %
Therefore, at the exit of the plasma, the beam envelope is assumed to be at waist having its transverse size increased due to its free expansion~\cite{Antici.2012}.} Assuming a sufficiently long propagation length in vacuum after the beam exits the plasma, \rev{one can express eq.~\eqref{eq. norm emittance with energy spread}} in terms of the drift length \rev{$L_{\mathrm{drift}}$} as derived in previous studies~\cite{Antici.2012, Migliorati.2013}:
\begin{equation}
    \epsilon_n^2 = \, \langle \gamma \rangle^2 \left( \sigma_E^2 \, \divRMS^{4} \, L_{\mathrm{drift}}^2 + \epsilon_{\rms}^2\right) \,.
    \label{eq. norm emittance with energy spread and divergence}
\end{equation}

For radio-frequency (RF) linear accelerators, where the energy spread $\sigma_E/E \to 0$, eq.~\eqref{eq. norm emittance with energy spread} reduces to the usual expression of $\epsilon_n \approx \, \langle \gamma \rangle \, \epsilon_{\rms}$. However, as noted in Refs.~\cite{Antici.2012, Migliorati.2013}, this approximation does not hold for laser-plasma accelerated (LWFA) electron beams, since the first term of the right-hand-side becomes much larger than the second term due to the large energy spread of the beam. \rev{Consequently, the normalized emittance becomes approximately $\epsilon_n^2 \approx \, \langle \gamma \rangle^2 \sigma_E^2 \, L_{\mathrm{drift}}^2 \left( \epsilon_{\rms}/\sigma_x \right)^4$, where the normalized emittance depends now on the electron beam geometric emittance and beam waist since $\divRMS = \epsilon_{\rms}/\sigma_x$.}

In addition, the geometric emittance represents the area encompassed by the trace space of the particle beam given by the position $x$, and the divergence $x^\prime$ of the particles within~\cite{Humphries.1990, Zhang.1996, Shanks.2009}. The area of the particle trace space is approximately the product of the source size $\sigma_x$ of the beam in its waist, and its divergence $\divRMS$. Hence, by measuring the geometric emittance of a beam and its divergence, it is possible to estimate its source size. Conversely, the emittance may be calculated by measuring both the divergence and the source size as in the laser interference method discussed below.

\subsection{The laser interference method}
In this section, we briefly review the theory of the laser interference method used in this work to estimate the emittance and source size of laser-accelerated electron beams with a particular emphasis on the robustness of the measurement process. 

\begin{figure}
\includegraphics[width=\columnwidth]{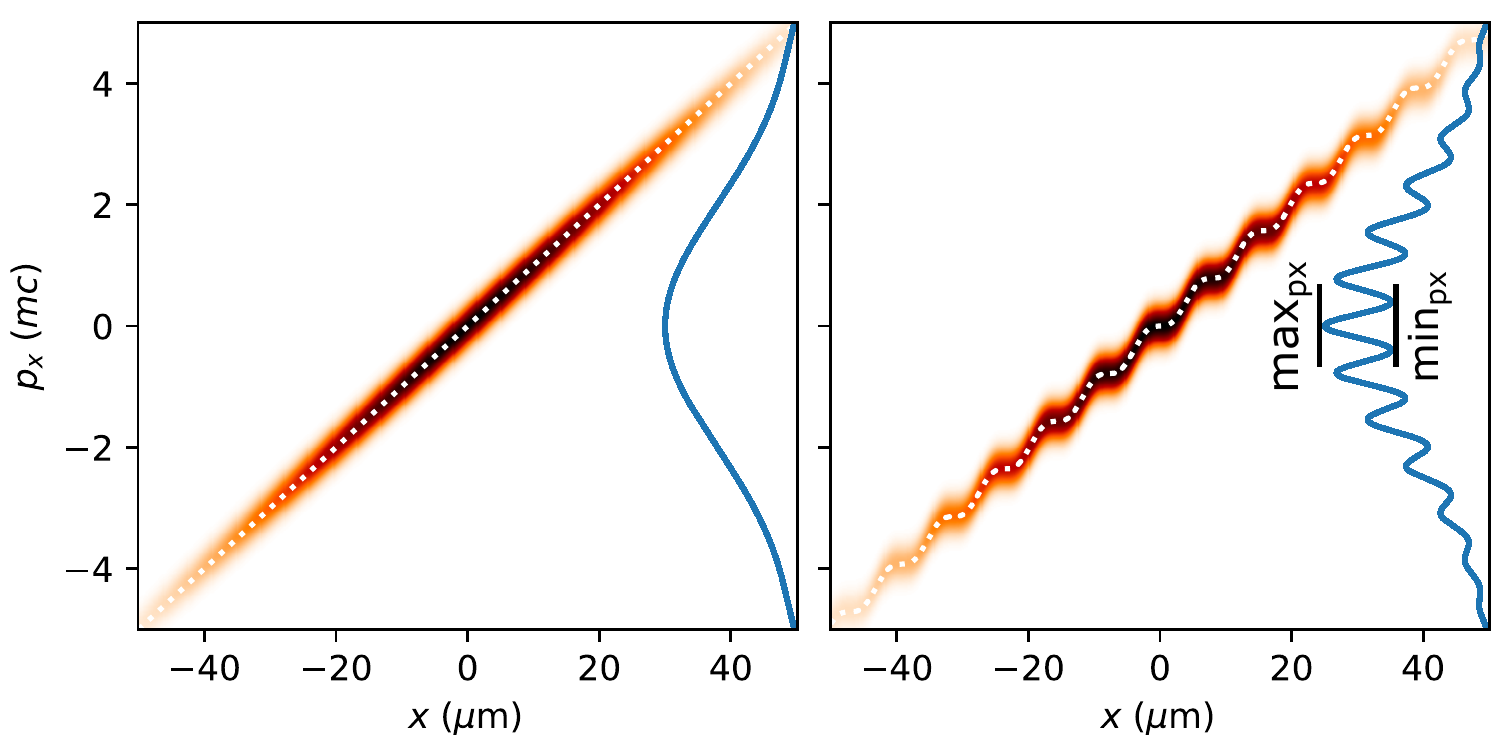}
\caption{Comparison between the phase space distributions $n(x,p_x)$ of the electron beam directly before (left) and after (right) its interaction with the laser interference pattern with intensity of $\kappa=1$ and wavevector of $k_G=0.8/\si{\um}$.  The initial electron beam parameters used for this simulation were: $\sigma_x=2$~\microm and $\sigma_p/m_e c=2$ and $L=10\,\si{\um}$ \rev{yielding in a ratio $\sigma_x/\lambda_G \approx 0.25$.} Blue curves show the integrated momentum distributions $N(p_x)$. \rev{The $\max_{p_x}$ and $\min_{p_x}$ of $\mathcal{F}(p_x, \,\pertpar)$ used for estimating the $R_T$ ratio as given in eq.~\eqref{eq. peak-to-valley ratio} are shown in the modulated signal in the left panel.}}
\label{fig. phase space ellipse}
\end{figure}

We take as the starting point of our consideration the beam waist of the electron beam (or the effective source at the output of the laser-wakefield accelerator prior to free expansion). We model the beam as uncorrelated Gaussian distributions with a spatial (rms) source size $\sigma_x$ and momentum rms width $\sigma_p$. The beam is then transported to the location of the laser grating, which we assume here as a free drift space of length $z_{\mathrm{drift}}$. \rev{This tilts the beam ellipse, which has a slope $dp_x/dx=1/L$, where $L = z_{\mathrm{drift}}/p_z$ represents the drift space parameter, where $p_{x, z}$ are the normalized momenta in x- and z- directions given as $p_{x, z} = \gamma \beta_{x, z}$ with $\gamma$ as the Lorentz factor of the particle.} The left panel of Fig.~\ref{fig. phase space ellipse} illustrates the rotation of the phase space ellipse due to the free drift of the particles for $L=10$~\microm. 

The intensity grating generated by the laser interference pattern pushes the electrons from the pattern's high-intensity peaks towards the low-intensity regions due to the ponderomotive force. This creates modulations in the density distribution of the electron bunch after the laser-electron interaction as,
\begin{multline} \label{eq:n-grating}
    n(x,\,p_x) = \rev{n_0} \exp \left[ - \frac{(x - L \, (p_x + \delta p_x))^2}{2 \sigma_x^2} \right] \, \\
    \times \exp \left[ - \frac{(p_x + \delta p_x)^2}{2 \sigma_p^2} \right] \, ,
\end{multline}
where $\delta p_x(x) = U \sin (k_G x + \varphi)$ is the periodic ponderomotive kick which depends on the grating wave vector $k_G$ and a phase offset $\varphi$ which will be set to zero in the following. The parameter $U = \max_x \left[ \langle F_\mathrm{pond}\rangle t_\mathrm{int} \right]$ parametrizes the strength of the ponderomotive scattering, and it depends on the maximum of the ponderomotive force $F_\mathrm{pond}$, and the interaction time $t_\mathrm{int}$ between the electron beam and the laser grating. According to eq.~\eqref{eq:n-grating} the modulations due to the laser grating have a periodicity of $\lambda_G = 2\pi/k_G = \lambda_L/(2 \cos \vartheta)$ in $x$ and a periodicity of $\lambda_G/L$ in $p_x$, where $\lambda_L$ is the wavelength of the laser used to produce the interference pattern with a crossing angle $\vartheta$.

To illustrate the impact of the laser grating on the phase space of an electron beam, consider the rotated phase space ellipse after a free drift as shown in the left panel of Fig.~\ref{fig. phase space ellipse}. Propagating this beam through a laser grating with intensity parameter $\kappa = 1$ and a wavevector of $k_G = 0.8$/\microm, the modulated phase space is presented in the right panel of the same Fig.~\ref{fig. phase space ellipse}. The parameter $\kappa$ represents the grating intensity in relation to the theoretical matched laser intensity. This parameter will be discussed in detail further below.

Such modulations in the phase space of the electron beam are detected via position-integrated transverse momentum distributions $N(p_x) = \int dx \, n(x, \,p_x)$. The strength of the modulation depends on the parameter $\pertpar = L\,U/\sigma_x$, such that $N(p_x) \to N(p_x, \, \pertpar)$. Note that the limit $\pertpar\to 0 $ corresponds to the absence of the laser grating and the electron momentum distribution is then reduced to $N(p_x, \,0) =  \sqrt{2\pi} \, \sigma_x \,e^{-p_x^2/2\sigma_p^2}$. As was shown first in Ref.~\cite{Seidel.2021}, measuring $N(p_x, \, \pertpar)$, in particular its modulation depth, allows to infer the electron beam source size $\sigma_x$, and hence the beam emittance.

To determine the source size of the electron beam, it is convenient to work with the ratio between the modulated and the non-modulated electron densities,
\begin{equation} \label{eq. ratio of spatial integrals}
    \mathcal{F}(p_x,\, \pertpar) = \frac{N(p_x, \, \pertpar)}{N(p_x, \, 0)} \,.
\end{equation}

Note that, for the case where the co-propagating laser is not present, i.e.~$\pertpar \to 0$, $\mathcal{F}(p_x,0) = 1$; there is no modulation of the electron beam.
With the reasonable assumption that $\sigma_p \ll \lambda_G/L$, the modulation depth function $\mathcal F$ can be approximately written as
\begin{align}
    \mathcal{F}(p_x,\, \pertpar) = \int_{-\infty}^\infty \frac{d\eta}{\sqrt{2\pi}} \exp \left[-\frac{1}{2}(\eta-\pertpar \sin\xi)^2
    \right] \,,
    \label{eq. ratio of spatial integrals 2}
\end{align}
where we introduced the normalized integration variable $\eta = (x - L \, p_x)/\sigma_x$ and $\xi = k_G \sigma_x  \eta + k_G  L p_x$. 

From the analysis of the experimental electron scattering data at a plane far from the laser interaction, the maxima ($P_{max}$) and minima ($P_{min}$) of the measured electron beam  distribution  $N(p_x)$ can be directly obtained. By comparing the experimental peak-to-valley ratio, $R_M = P_{max} / P_{min}$, with the corresponding theoretical expression $R_M=R_T$, with
\begin{align}
    R_T = \frac{\max_{p_x} \, \mathcal{F}(p_x, \,\pertpar)}{\min_{p_x} \, \mathcal{F}(p_x, \,\pertpar)} \,,
    \label{eq. peak-to-valley ratio}
\end{align}
we are able to extract the electron beam source size $\sigma_x$.
For instance, for small values of $\pertpar$ and $\kappa = k_G L U$ the integral in eq.~\eqref{eq. ratio of spatial integrals 2} can be solved perturbatively yielding 
$R_T \simeq 1 + 2 \kappa e^{-k_G^2 \sigma_x^2/2}$. If $\kappa$ is not small, eq.~\eqref{eq. peak-to-valley ratio} can be solved numerically for the electron beam source sizes $\sigma_x$, given the laser grating wave vector $k_G = 2\pi/\lambda_G$ (with laser grating wavelength $\lambda_G$), drift length parameter $L$, and grating strength $U$ are known.

\subsection{Robustness of the Method}

\subsubsection{Optical Grating Strength}

While the parameters $\lambda_G$ and $L$ are relatively straightforward to determine experimentally with good accuracy, the grating strength $U$, or $\kappa$, can be much more challenging to measure precisely. The meaning of the parameter $\kappa$ is the ratio of the laser grating intensity $\IntLaser$ to the matched intensity $\mathcal  I_0$ \cite{Seidel.2021}, $\kappa = \IntLaser/\IntLaser_0$, for which the modulated phase space ellipse exhibits horizontal tangents after its interaction with the ring-like laser beam, see Fig.~\ref{fig. phase space ellipse}. The condition when this is satisfied is determined by $k_G L U  = 1$, which implies for the matched laser intensity
\begin{equation}
\IntLaser_0[10^{18}\,\si{\watt}/\si{\cm}^2] = 0.277 \times \lambda_L^2[\si{\um}] \frac{\gamma^2 \lambda_G^2}{z_\mathrm{drift} \: c \: t_\mathrm{int}} \,,
    \label{eq. I0 for tangent at phase-space}
\end{equation}
where $\gamma$ is the Lorentz-factor of the electron beam, and $c$ is the speed of light in vacuum. 

\begin{figure}
    \centering
    \includegraphics[width=\columnwidth]{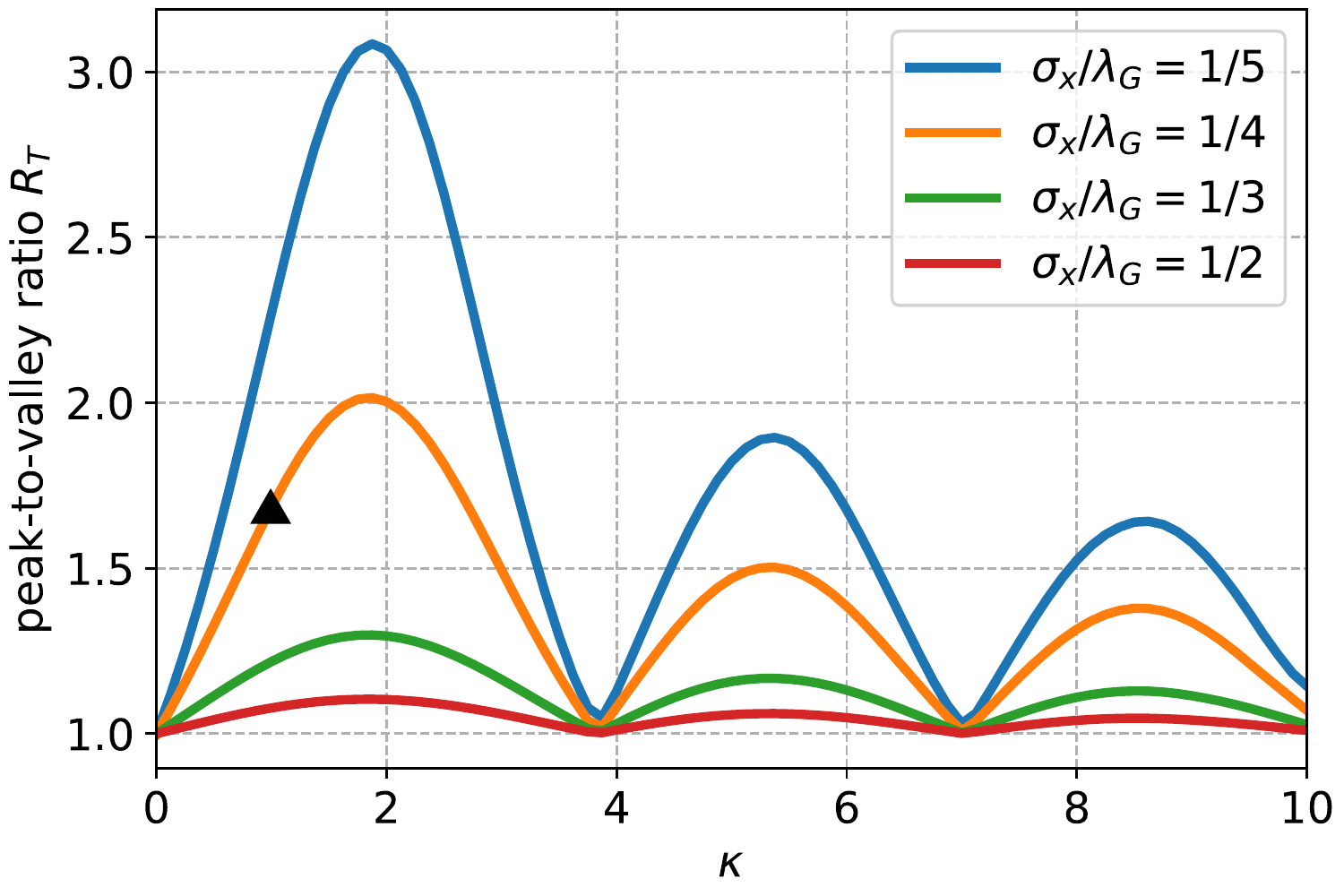}
    \caption{Theoretical calculation of the peak-to-valley ratio $R$ of the modulation as a function of $\kappa$ for various source sizes $\sigma_x$. \rev{The triangle marker represents expected peak-to-valley ratio $R_T$ for the simulated modulation in the beam distribution shown in Fig.~\ref{fig. phase space ellipse}.}}
    \label{fig. pvr with respect to kappa}
\end{figure}

In Figure~\ref{fig. pvr with respect to kappa} we show a numerical calculation for the peak-to-valley ratio $R_T$, eq.~\eqref{eq. peak-to-valley ratio}, as a function of $\kappa$. The different curves are for various values of the dimensionless ratio $\sigma_x/\lambda_G$. For small $\kappa$, the peak-to-valley ratio rapidly increases above unity and then peaks at around $\kappa\approx 2$ for the depicted $\sigma_x/\lambda_G$. For even larger $\kappa$ the value of $R_T$ drops again and then oscillates.

\begin{figure}
    \centering
    \includegraphics[width=\columnwidth]{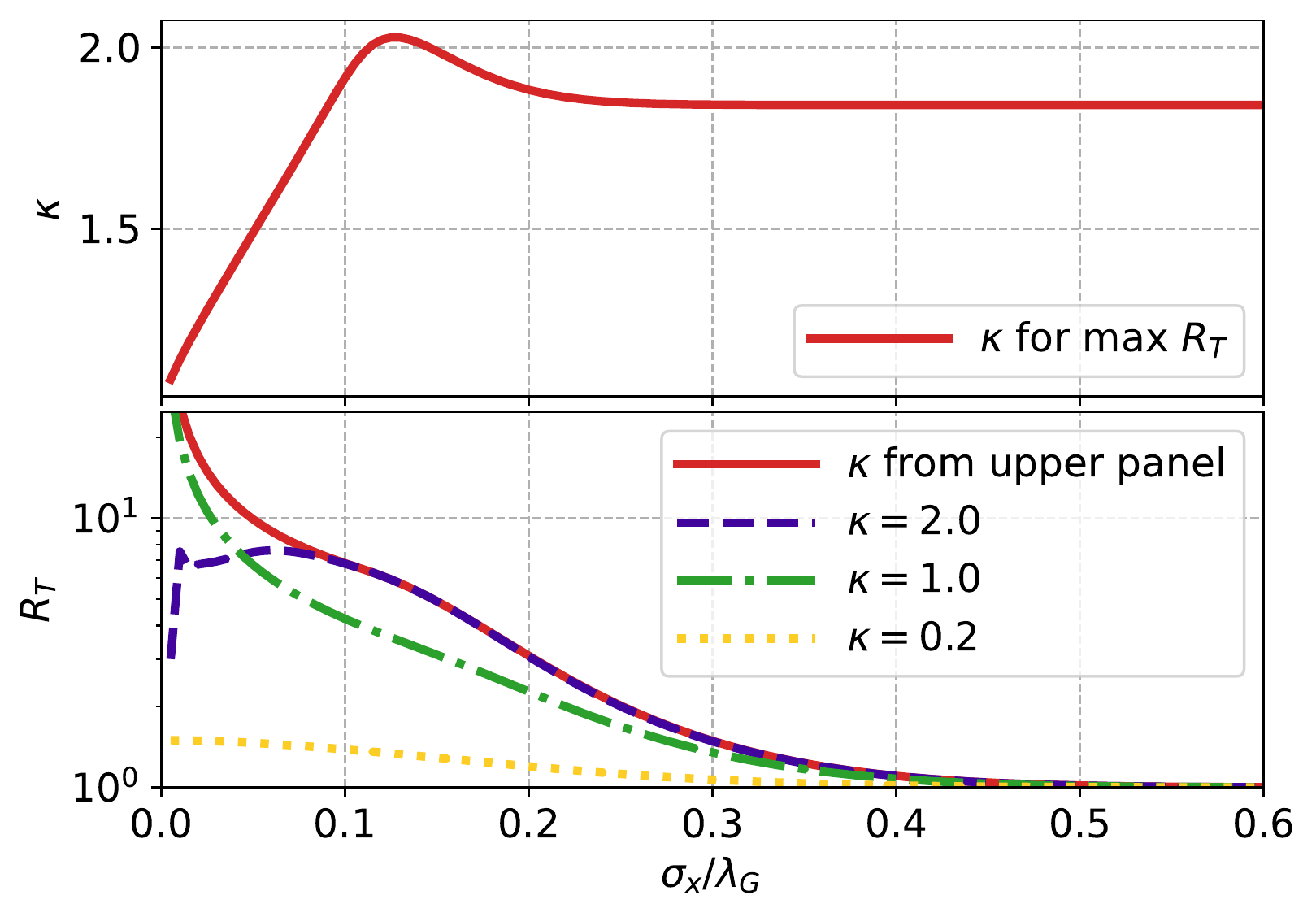}
    \caption{Upper panel: Grating intensity parameter $\kappa$ that maximizes the peak-to-valley ratio $R_T$ of the modulated electron beam as function of various source sizes (solid red curve). 
    Lower panel: Maximum $R_T$ achievable for given source size for various $\kappa$. In addition to the $\kappa$ from the upper panel, the purple dashed curves correspond to the case of $\kappa=2$, green is for $\kappa=1$, and the yellow dotted curve corresponds to $\kappa=0.2$. The values of $R_T$ at $\kappa=2$ and the maximum are extremely close except for very small source sizes.}
    \label{fig:optimal_kappa}
\end{figure}

In Figure~\ref{fig:optimal_kappa}, we show in the upper panel the value of $\kappa$ that yields the largest peak-to-valley ratio of the modulated electron beam at given source size $\sigma_x/\lambda_G$. We find that only for very small $\sigma_x/\lambda_G$ the most sensitive value of $\kappa$ approaches one. However, for all experimentally relevant values $\sigma_x/\lambda_G>0.1$ we find $\kappa\approx2$ gives the highest modulation on the electron beam. In the lower panel of Fig.~\ref{fig:optimal_kappa} we also show the maximum achievable $R_T$ (i.e.~for the $\kappa$ from the upper panel that maximize the modulation) for given source size $\sigma_x/\lambda_G$. Additionally, we plot the peak-to-valley ratios for $\kappa=0.2, 1, 2$, showing that $\kappa=2$ yields almost the same maximum $R_T$ for all but very small source sizes. For the latter, $\kappa=1$ is an optimal choice due to its high sensitivity for small source sizes.

\begin{figure}
    \centering\includegraphics[width=\columnwidth]{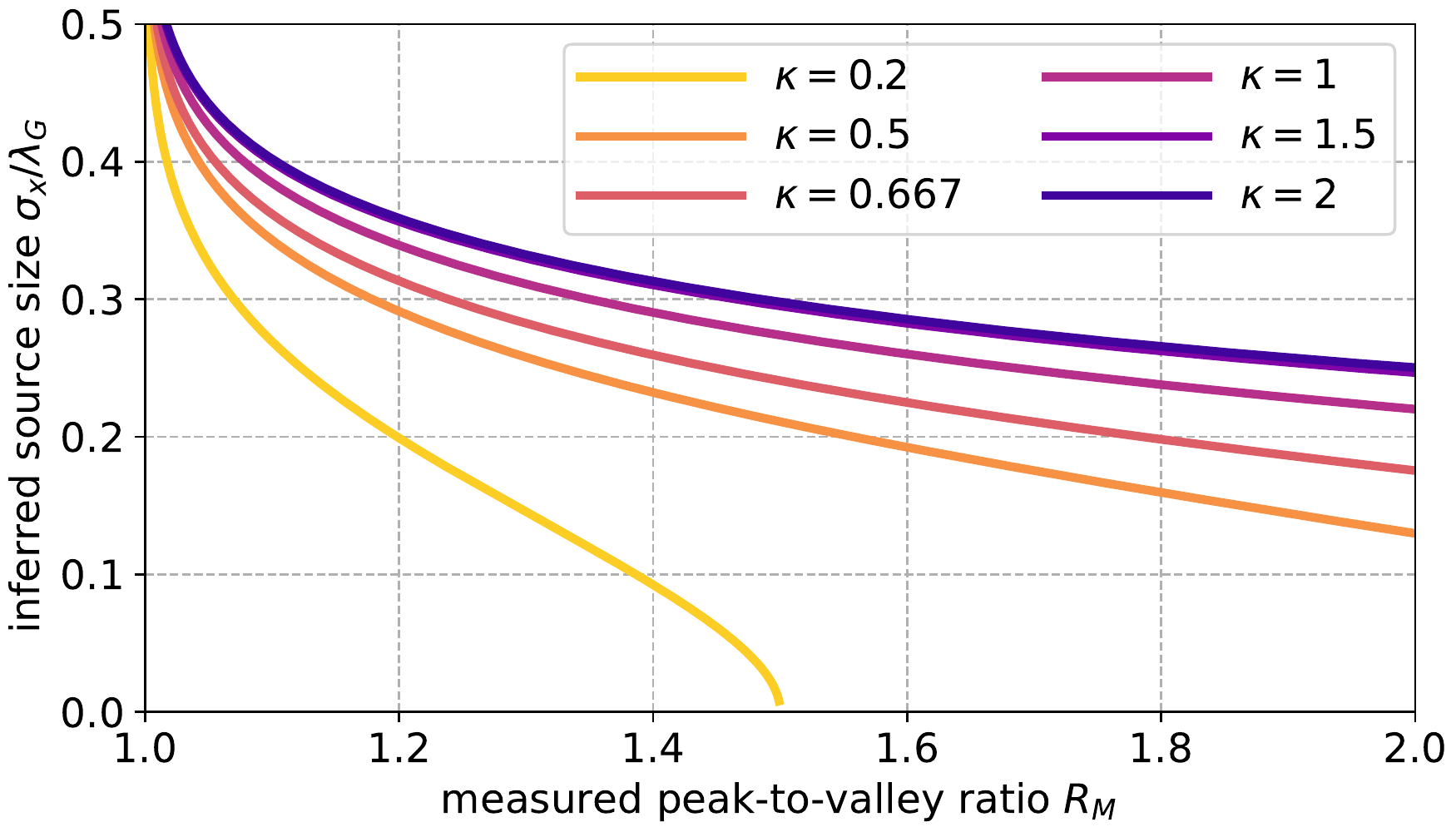}
    \caption{Influence of the laser grating strength $\kappa$ on the source size inference for given peak-to-valley ratio $R_M$.}
    \label{fig:kappa effect}
\end{figure}

The influence of the grating strength $\kappa$ on the inferred value of $\sigma_x$ for given peak-to-valley ratio can be more clearly seen in Figure~\ref{fig:kappa effect}. In this plot, the various curves represent how a measured value of $R_M$ maps to a source size $\sigma_x$ for various values of $\kappa$. Clearly, the curve for $\kappa = 2$ gives an upper limit for the possible source for any given experimentally determined peak-to-valley ratio.
The most precise estimate is possible for a known $\kappa$, however analysing any given data set under the assumption provides an upper bound for the true source size $\sigma_x$, limiting the importance of accurate knowledge of the grating strength. In situ determination of $\kappa$ is of course possible by scanning the laser energy and deriving $\kappa$ from the predicted $R_T(\kappa)$ shown in Figure \ref{fig. pvr with respect to kappa}.
This approach is applicable for all $\sigma_x > 0.1\lambda_G$.

\subsubsection{Influence of background noise}

Generally,  background noise or background  originating from electrons that may not have interacted with the laser grating can affect the peak-to-valley ratio $R_T$. This is particularly important for electron beams that are not monoenergetic such as the ones in the test experiment described below. Lower energy particles slip backwards compared to the main spectral peak and hence may interact less with the laser grating beam. Consequently, they do not contribute to the modulation depth but rather manifest as background signal in our measurements, thus effectively reducing the observed peak-to-valley ratio.

Let us assume that a fraction $\nu_B$ of all electrons constitute a uniform level of the background signal across the modulations. Then, the observed peak to valley ratios are related to the theoretical modulation depth $\mathcal F$ via the weighted sums
$P_M = [(1-\nu_B) \max \mathcal F + \nu_B]N(0)$ and
$V_M = [(1-\nu_B) \min \mathcal F + \nu_B]N(0)$, where $N(0)$ is the total unmodulated signal.
Thus, we obtain the modified peak-to-valley ratio
\begin{align}
    \frac{P_M}{V_M} = \frac{\max \mathcal F + \mathcal B}{\min \mathcal F + \mathcal B} \,,
\end{align}
where $\mathcal B = \nu_B/(1-\nu_B)$. 

\begin{figure}
    \centering\includegraphics[width=\columnwidth]{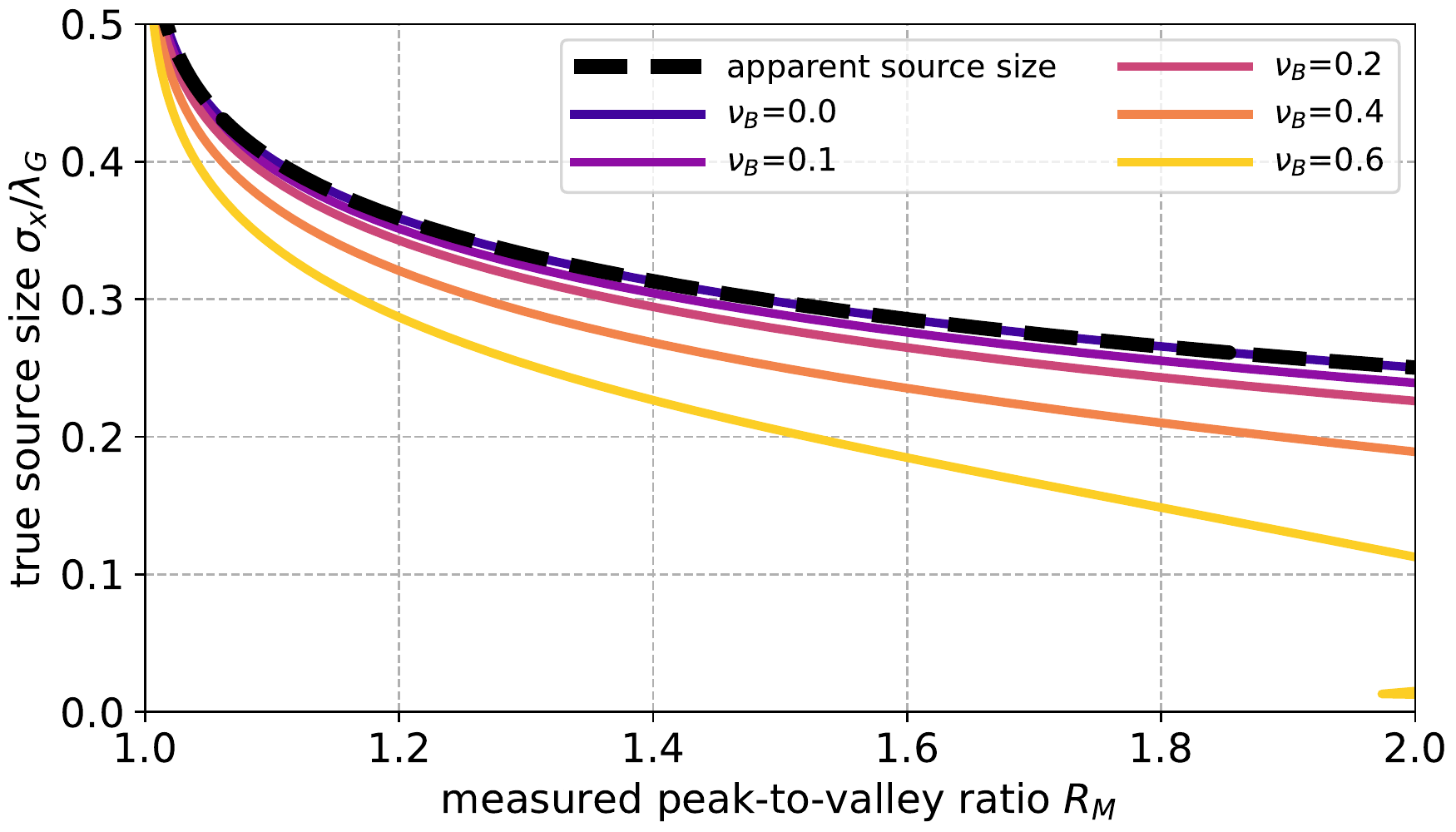}
    \caption{Effects of the background noise on the estimation of the source size of the electron beam calculated for $\kappa=2$. When setting $\mathcal B=0$, i.e., $\nu_B = 0$, no background noise is present, the apparent source size value is maximum representing an upper limit of its true value.}
    \label{fig:background noise effect}
\end{figure}

To closely examine the impact of background noise, Fig.~\ref{fig:background noise effect} illustrates the influence of different background noise ratios when estimating the source size. The results demonstrate that neglecting background in the data analysis, i.e., setting $\nu_B = 0$ and $\mathcal B=0$, leads to an apparent value of $\sigma_x$ that represents an upper limit of its true source size, represented in Fig~\ref{fig:background noise effect} by the dashed black line.

\subsubsection{Effects of large electron beam energy spread} 
The theoretical framework developed for estimating the source size using the laser grating method, as presented earlier, considers specifically monoenergetic electron beams.

To assess the impact of broad energy bandwidth beams, we conducted Particle-in-Cell (PIC) simulations with parameters closely resembling those of the electron beam and the laser grating strength employed in the experiment described in this work. The results of these simulations with large energy spread electron beams are given in Appendix~\ref{appendix density contribution for different energies}. From results of these PIC simulations, the modulated signals of the electron beam do not present any significant differences for various energy spreads. 

Consequently, employing the monoenergetic theory, as previously detailed in this section, for an electron beam that is, in reality, broadband, is completely valid to infer the source size and emittance of laser-accelerated electron beams. 

We note therefore that over a very wide range of conditions with regards to electron beam spectrum, background levels or laser grating strength the method is robust with regards to determining an upper limit for the source size. Very tight upper bounds can be determined in the limit of small laser grating parameters $\lambda_G$.

\section{Experimental Setup}
\label{sec experimental setup}

\begin{figure*}[t]
	\centering
	\includegraphics[width=0.95\linewidth]{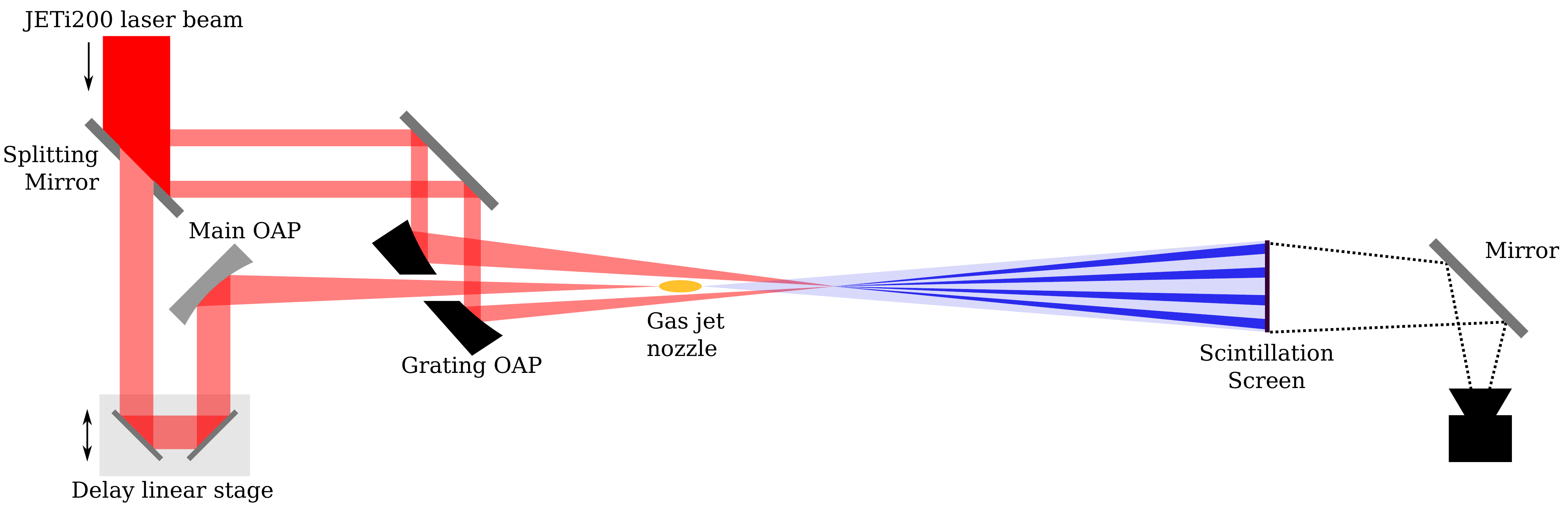}
	\caption{Experimental layout for laser interference method test. The laser interference pattern was generated by focusing the ring laser beam by an off-axis parabolic mirror (grating OAP). The electron beam (blue) was modulated after the interaction with the laser grating and then propagated to a scintillation screen, which was imaged onto a CCD camera.}
	\label{fig. experimental setup up}
\end{figure*}

To demonstrate the principle of emittance and source size measurements using laser-gratings, we conducted an experiment at the Helmholtz Institute Jena using the JETi200 laser system (7.2~J, 800~nm center wavelength, 23~fs pulse duration). The experimental setup is shown in Fig.~\ref{fig. experimental setup up}. During the experiment, the JETi200 laser beam was split into two beams: the central beam, with a diameter of 60~mm, which is used to accelerate electron beams via the LWFA, and a fraction the remaining outer ring which is employed to create the grating pattern and modulate the electron beam.

The central beam was focused by an off-axis parabolic mirror (OAP) of f-number~=~16.7 to an elliptical spot with major axis of $(23.7 \pm 1.8)~\si{\um}$ (FWHM), corresponding to an intensity of $7\times 10^{18}$~W/cm$^2$, into a supersonic gas jet (mixture of 95\% He and 5\% N$_2$) generating an underdense plasma with an electron density of $1.1\times10^{19}$~cm$^{-3}$ in the plateau region. For these parameters, injection via ionization in the plasma bubble is the dominant injection mechanism~\cite{Kuschel.2018}.

\begin{figure}
	\centering
    \includegraphics[width=0.7\linewidth]{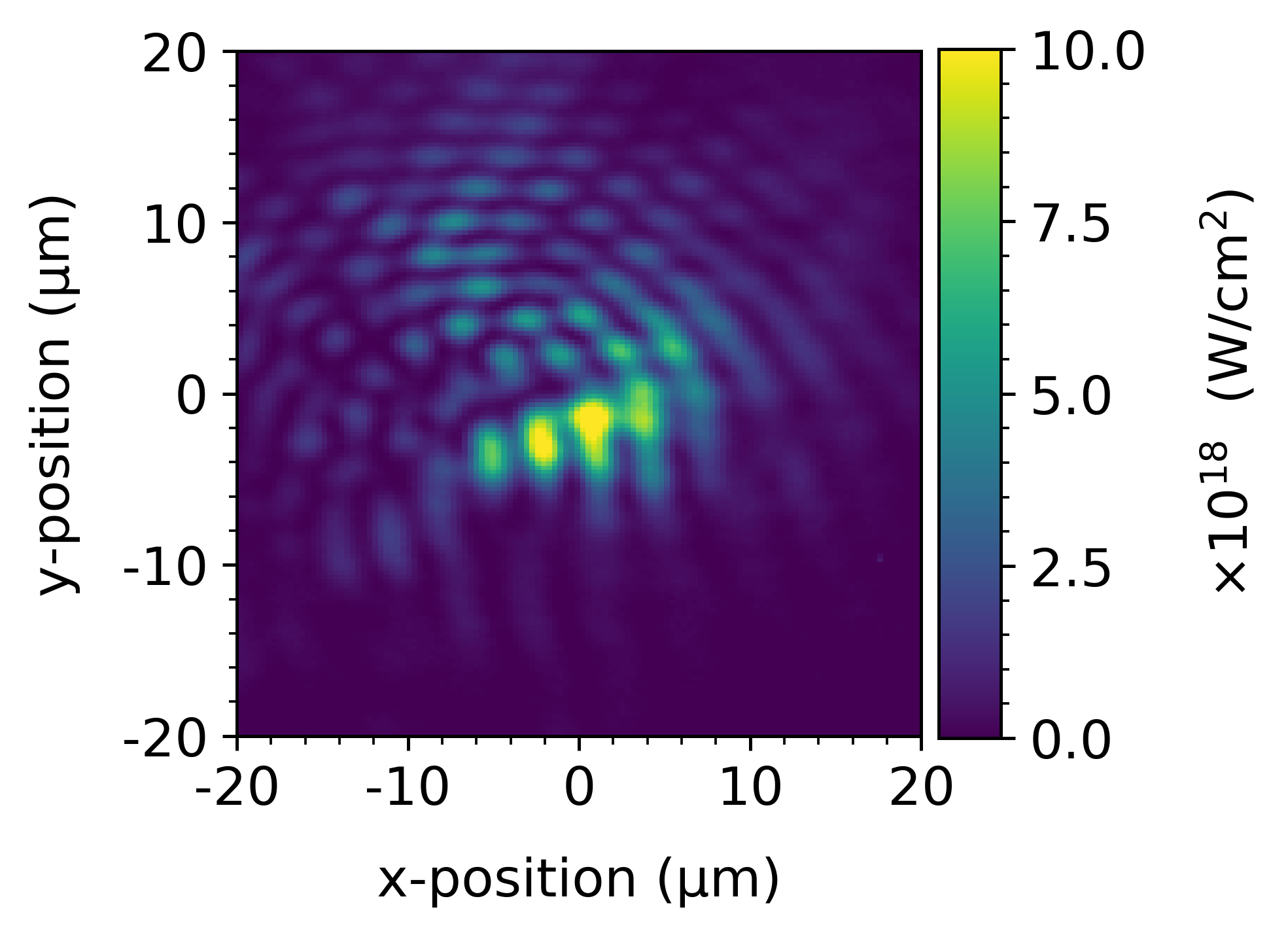}
	\caption{Focal spot of the ring laser. An interference pattern with a mean distance between peaks of $\lambda_G = (4.2~\pm~0.1)$~\microm and peak intensity of $\approx 10^{19}$~W/cm$^2$ is observed. The interference pattern shown here is used to modulate the electron beam transverse phase-space for obtaining the source size and emittance of the electron beam.}
	\label{fig. pick-off laser beam}
\end{figure}

The ring laser beam underwent focusing using an off-axis parabolic mirror (grating OAP) with a focal length of 40 cm. This focusing occurred approximately \rev{$z_{\mathrm{drift}} = 34.4~\si{\mm}$} downstream of the gas jet nozzle, where the produced interference pattern interacted with the electron beam. The resulting laser interference pattern, generated by the focused ring-shaped beam, is depicted in Fig.~\ref{fig. pick-off laser beam}.

It is important to note that due to limitations in the experimental setup, the focus spot of the ring laser beam is not diffraction-limited. This limitation arises from the wavefront optimization which was set for the LWFA beam. Consequently, wavefront distortions remained and occurred at the edges of the splitting mirror and the diamond turned parabola resulting in non-ideal focusing visible in Fig.~\ref{fig. pick-off laser beam}. \rev{As consequence of wavefront distortions, the ring laser displays several vertical and concentric fringes. During data evaluation, we focus on the region where the vertical fringes interact with the electron beam. In this area, the vertical fringes induce also modulations in the electron beam density in the same direction, along the y-axis. Consequently, we are able to evaluate the beam waist perpendicular to the fringes, denoted as $\sigma_x$. On the other hand, to measure the vertical beam waist and emittance, the interference pattern of the laser beam must be oriented horizontally, i.e., along the x-direction.}

For our experimental conditions the focus of the ring-like laser beam (focal diameter on the order $\mathcal{O} \sim 20~\si{\um}$) was significantly smaller than the electron beam ($\mathcal{O} \sim 180~\si{\um}$). The electron beam was therefore always interacting with the full range of intensities up to the maximum value, implying that the strongest modulation visible will automatically correspond to $\kappa = 2$.
 
Performing a vertical integration around the highest peak intensity in the y-direction of the focal spot with the interference pattern, we obtain a mean peak-to-peak distance of $\lambda_G = (4.2\pm0.1)$~\microm between the fringes in the ring beam, leading to a grating wavevector $k_G = 2\pi/\lambda_G \approx (1.5 \pm 0.1)/\si{\um}$.

The electron beam is detected by a YAG:Ce scintillation screen of 100~\microm thickness \rev{placed about $L_{\mathrm{drift}} \approx 1415~\si{\mm}$ downstream of the focal spot of the ring laser.} The scintillator converted the electron beam profile into a similar radiation signal at a wavelength of 550~nm. The screen was imaged with a resolution of 18.5~\microm by an Andor Marana camera, model~4.2B-6 (16-bit, quantum efficiency of approximately 95\% at 550~nm)~\cite{OxfordInstruments.2022}.
\section{Experimental Results and Discussions}

The test electron beams exhibited a wide energy spectrum (from 40 MeV to 120 MeV) with \rev{an average charge of $(5.6 \pm 0.7)$~pC per bunch}. The weighted mean of the electron beam energy is calculated to be approximately 73~MeV, which leads to a weighted mean Lorentz factor of $\langle \gamma \rangle~\approx~143$, which was used later to evaluate the normalized emittance as given in eq.~\eqref{eq. norm emittance with energy spread}.

\begin{figure}
	\centering
	\includegraphics[width=0.95\linewidth]{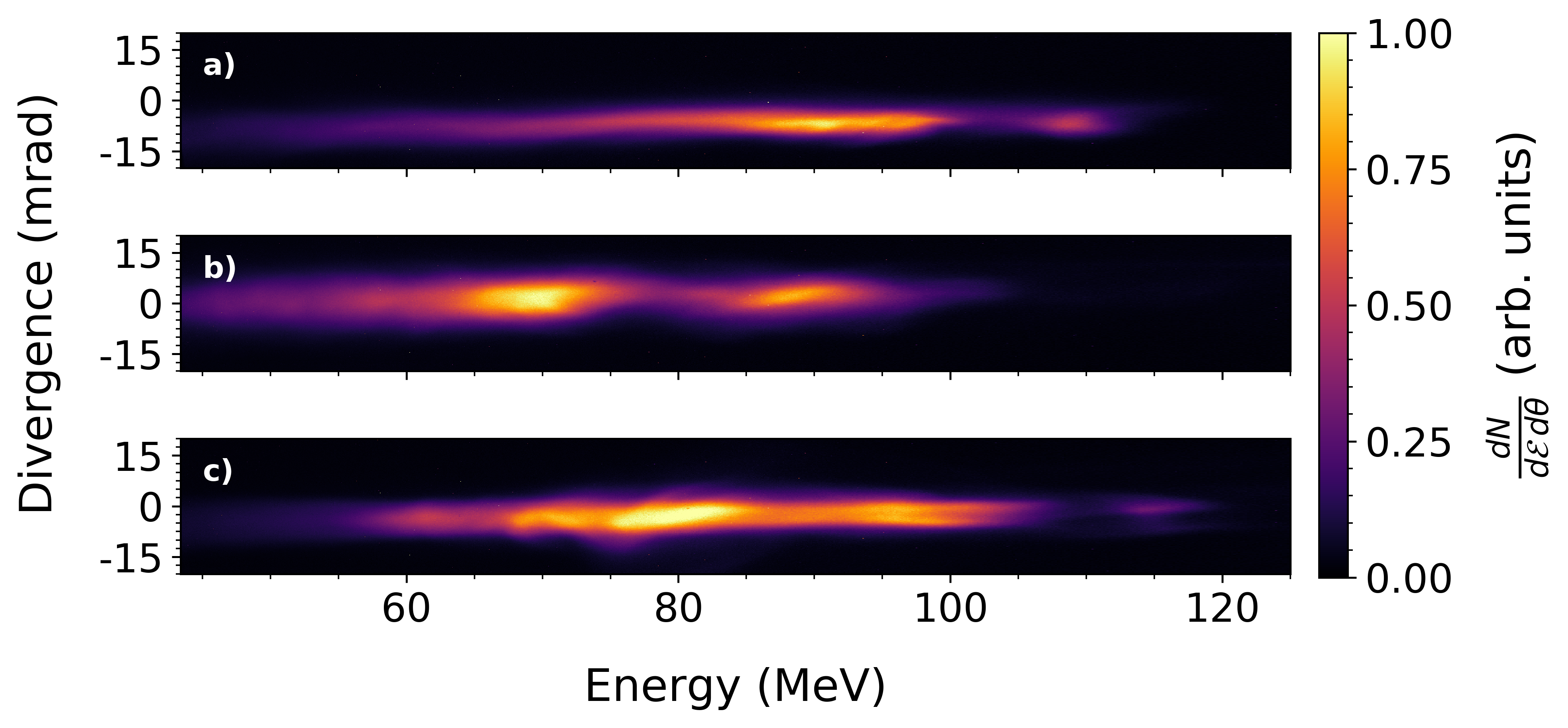}
	\caption{Electron beam energy spectra taken from three shots with the same experimental settings.}
	\label{fig. emittance energy spectrum}
\end{figure}

The energy spectra of the three representative shots taken during the experiment are shown in Fig.~\ref{fig. emittance energy spectrum}. The spectral diagnostic installed on the experiment was capable of characterizing electrons with kinetic energies above approximately 50~MeV, which was sufficient to determine the significant energy spread of the beam. A low energy (few-MeV) spectral component  is also expected to be generated due to the injection mechanism employed on the laser wakefield accelerator~\cite{Pak.2010, McGuffey.2010}. These particles have a large angular spread and do not contribute significantly to our measurements, only as background signal on the measurements which were taken into account during the evaluation process. From the energy spectrometer measurements, a weighted average energy spread of $\sigma_E/E = (27.3 \pm 4.8)$\% was obtained. \rev{In addition, an average rms divergence of unmodulated electron beams of $\divRMS\approx(2.6\pm0.4)$~mrad was evaluated for 130 shots. To experimentally determine the rms divergence $\divRMS$, the ring laser was switched off (blocked) to avoid any modulation on the electron beam to be created. With the ring laser switched off, the beam profile was recorded on the scintillation screen, and the rms divergence was retrieved through Gaussian fitting on each recorded shot. We assumed that the divergence remained constant throughout the entire experimental conditions.}

\subsection{Source size and emittance evaluation}
\label{sec. experimental results laser interference}

\begin{figure}
	\centering
	\includegraphics[width=\linewidth]{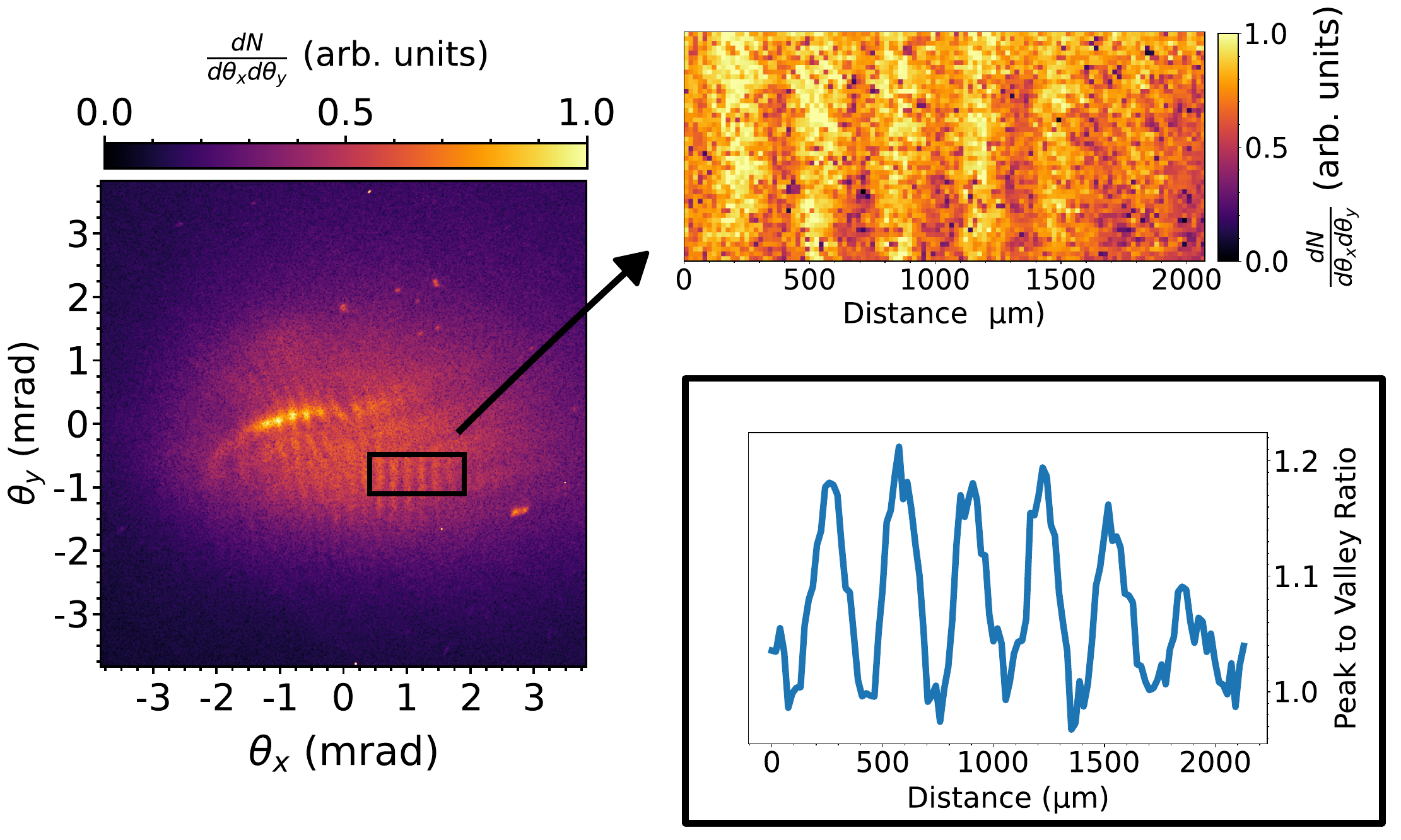}
	\caption{Modulated electron beam due to the interaction with the interference fringes of the ring laser beam at the focus. (Left panel) Electron beam was imaged with a YAG:Ce scintillation screen and has a mean peak-to-peak distance between the fringes of about (330.1~$\pm$~6.6)~\microm. (Upper right) Zoom image at the region of interest where the modulations were taken for data analysis. (Lower right) Integrated peak-to-valley ratio modulation used for the source size analysis using the laser grating method. \rev{The divergence in the x-direction $\divRMS$ is utilized for inferring the beam waist using the laser-grating method. The vertical divergence, i.e., in the y-direction, is denoted as $\theta_y$.}}
	\label{fig. modulated ebeam and background subtracted}
\end{figure}

To evaluate the source size and emittance of the electron beams generated during the experiment, we proceeded with the procedure explained in section~\ref{sec experimental setup}. As soon as both, spatial and temporal, overlaps between the electron beam and ring-like laser pulse were achieved, the modulated electron beam propagated towards the scintillation screen. The left panel in Fig.~\ref{fig. modulated ebeam and background subtracted} shows an example of the modulated electron beam on the screen. 

An integration along the fringes within the region-of-interest (upper right panel in Fig.~\ref{fig. modulated ebeam and background subtracted}) was performed to obtain the baseline which represents the unmodulated electron beam $N(p_x, \, 0)$. 
\rev{The region-of-interest (ROI) for the data analysis was chosen where vertical fringes were observed, allowing us to assess the horizontal beam waist $\sigma_x$, and subsequently, the horizontal geometric emittance ($\epsilon_\mathrm{rms} = \epsilon_x$). Additionally, as long as the fringes are vertically oriented, the inferred values of $\sigma_x$ should be independent of the selected ROI, as all experimental parameters remain constant, especially the grating wavelength which is crucial for the inference of the emittance later.}

The peak-to-valley of the signal was then determined to evaluate the source size of the electron beam. In the lower right panel in Fig.~\ref{fig. modulated ebeam and background subtracted} is shown an example of an integrated signal of a modulated electron beam recorded in the experiment \rev{(upper right panel of Fig.~\ref{fig. modulated ebeam and background subtracted})}. \rev{The integrated signal represents the modulation on the electron beam as described in eq.~\eqref{eq. ratio of spatial integrals}, allowing us to extract the maximum and minimum of the density modulation to evaluate the beam waist as in eq.~\eqref{eq. peak-to-valley ratio}.} By analysing a total of 184 shots, an average peak-to-valley ratio of about $(1.09\pm0.04)$ for our experimental conditions was obtained.

The experimental peak intensity in the interference pattern, exceeded the  the critical laser intensity, as defined in eq.~\eqref{eq. I0 for tangent at phase-space} significantly. For our conditions the intensity corresponding to $\kappa=2$ is evaluated as $\IntLaser_{0} = 6.4\times10^{18}\, \si{\watt \cm^{-2}} / t_\mathrm{int}[\si{\fs}] \approx 1.1\times10^{16}\, \si{\watt \per\cm^{2}}$, for an interaction time of $t_\mathrm{int} = 582~\si{\fs}$. The interaction time was determined using the geometry of the experiment. For details on the calculation of the interaction time, please see Appendix~\ref{appendix interaction time}.

\begin{figure}
    \centering
    \includegraphics[width=\columnwidth]{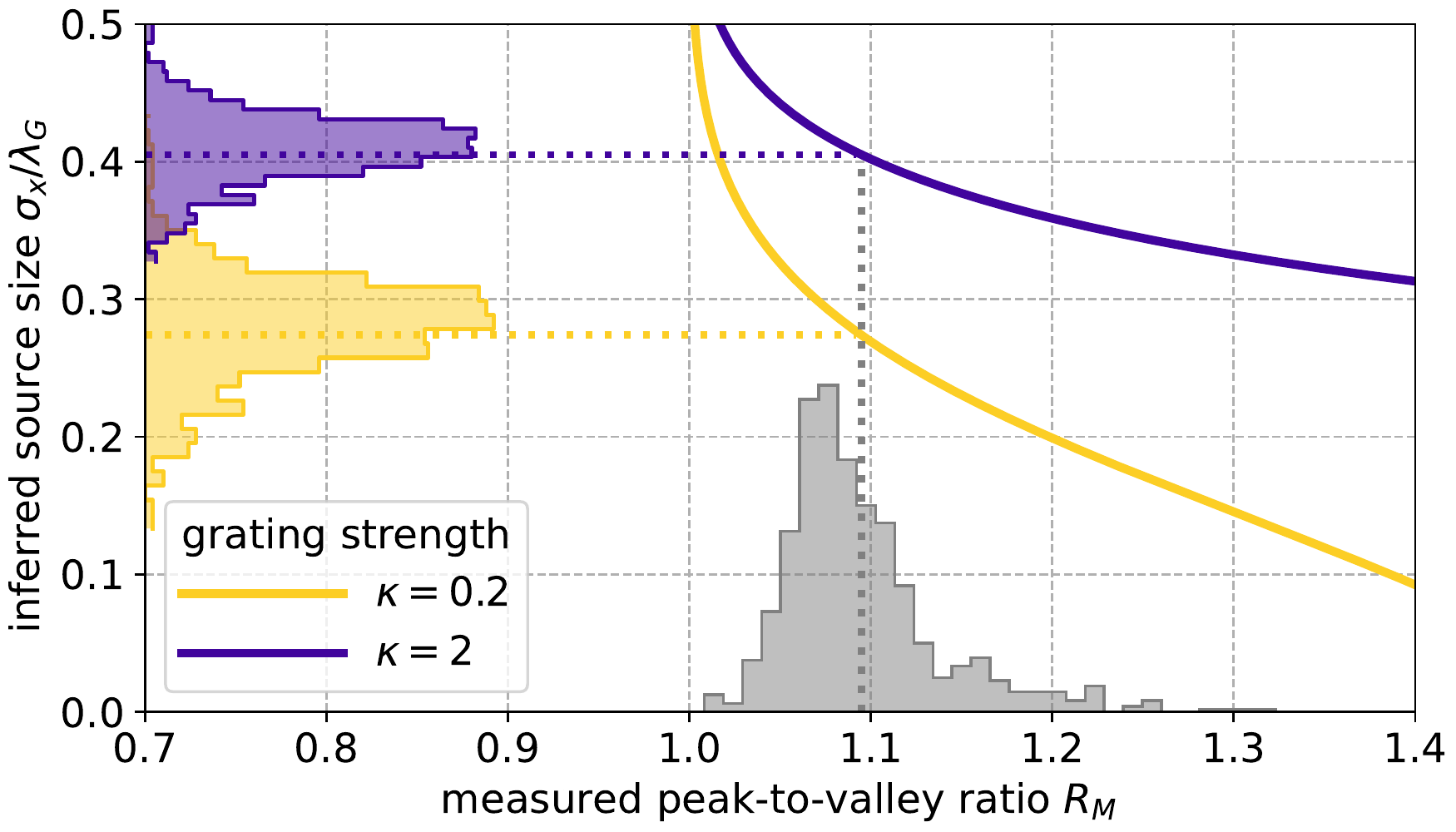}
    \caption{Histogram transformation of the measured peak-to-valley ratios of 184~shots (grey histogram on bottom axis) into inferred electron beam source size distributions at different grating strengths $\kappa$ (histograms on left axis). The purple curve and histogram correspond to the absolute upper limit for the inferred $\sigma_x$ (see discussion in text), while yellow corresponds to the measured grating strength of $\kappa=0.2$.}
    \label{fig:histograms}
\end{figure}

As the larger electron beam sampled the full range of laser grating intensities,  we  take the maximum modulation depth  to have been generated by the optimal intensity, i.e. we estimate the electron beam's source size and its geometrical emittance by taking the strongest modulation visible to correspond to $\kappa=2$.  As depicted in Fig.~\ref{fig:histograms}, we observe that the inferred source size is constrained by this assumption  as detailed explained in section \ref{theory laser interference method}. 

Assuming $\kappa =2$, the root-mean-square (rms) source size of the LWFA electron beam is $\sigma_x \approx (1.7\pm0.2)~\si{\um}$, with a corresponding geometric emittance of $\epsilon_{\rms} \approx (4.4\pm0.9)\times10^{-3}~\pi$~mm~mrad. 

To illustrate the weak dependence on precise value of $\kappa$ for our parameters we also include a simulation using a smaller values of the grating strength parameter $\kappa = 0.2$. Even with this much lower interaction strength the resulting source size estimate would only shift to $(1.2\pm0.4)$~\microm, and a geometric emittance of approximately $(3.1\pm1.2)\times10^{-3}~\pi$~mm~mrad. Hence, it is noteworthy that despite the order of magnitude difference in intensity between both emittance estimations with $\kappa = 2$ and 0.2, the source size and geometric emittance exhibit low sensitivity to $\kappa$.
This shows that  source size measurements depend only weakly on the precise the grating strength for small modulation depths $R_M$. The source size derived from the laser grating experiment aligns with reported values in the literature for LWFA electron beams, typically in the range of a few-micrometers, as reported in Refs.~\cite{Corde.2011, Schnell.2012, Plateau.2012}.
We note that in-situ calibration of $\kappa$ would be possible in experimental configurations where the laser grating pulse energy can be continuously varied and the pulse energy corresponding to $\kappa =2$ determined by comparison with the scaling shown in Fig.~\ref{fig. pvr with respect to kappa}. 

The measurement of very small source sizes will react most sensitively to changing the grating wavelength $\lambda_G$ and that obtaining strongly modulate electron beams with high peak to valley is unique to beams with very small ($\ll~\si{\um}$) sources sizes 

\rev{For completeness, the normalized emittance, assuming $\kappa = 2$ and taking into account the electron energy spread obtained during the experiment with a drift length $L_{\mathrm{drift}} = 1415~\si{\mm}$, is calculated as $\epsilon_n \approx 380~\pi~\si{\mm\mrad}$. This value simply highlights the fact that the normalized emittance is primarily influenced by the  energy spread for broad band electron beams.} Measurement of the normalized emittance for broadband beams is therefore best performed in conjunction with a dipole magnet  to provide spectral resolution.  Finally, we note that a direct comparison of the laser-grating method with   traditional methods such as the pepper-pot highlights the enhanced accuracy of the laser-grating method. For our experimental conditions, the pepper-pot method returns a geometric emittance of approximately $43\times10^{-3}$~$\pi$~mm~mrad, as detailed in our publication in Ref.~\cite{emittance-paper}, far above the literature values for LWFA beams obtained with quadrupole scans, due to the limitations of the pepper pot method for beams with small source sizes.

The source size derived from the laser grating experiment aligns with reported values in the literature for LWFA electron beams, typically in the range of a few-micrometers, as reported in Refs.~\cite{Corde.2011, Schnell.2012, Plateau.2012}.  Our work shows that precise measurements of $\sigma_x$ and tight upper bounds for electron beams with extremely small source sizes are possible using the laser interference method, in particular if the grating strength (ideally at $\kappa = 1$) is characterised in-situ  and background noise is kept to a minimum and the smallest appropriate $\lambda_G$ is chosen.

\section{Conclusions}

Here we presented the first experimental  demonstration of the laser grating method for emittance and source size measurements. The  source size and emittance for laser wakefield accelerator (LWFA) electron beam is determined to be  $(1.7\pm0.2)$~\microm, and the geometric emittance is  $(4.4\pm0.9)\times10^{-3}~\pi$~mm~mrad. This corresponds well to published values~\cite{Corde.2011, Schnell.2012, Plateau.2012} for comparable LWFA accelerators e.g. using quadrupole scans. 
We detail the scaling of the method with respect to grating strength and background noise and conclude that tight upper bounds can be determined and small source sizes accurately measured.  This is true even if there are uncertainties in the  grating intensity and in environments with low signal to noise ratio. Hence the laser grating method is a robust and sensitive approach to characterising electron beams with very low emittance and  small source size in single shots.

\section{Acknowledgements}

The authors thank G.~Schäfer for operating the JETi200-
laser system. This research was funded by the Federal Ministry of Education and Research of Germany (BMBF) in the Verbundforschungsframework (Project No. 05K19SJA). F.~C.~S. and M.~Z. thank the funding by the Deutsche Forschungsgemeinschaft (DFG) under Project No.~416708866 within the Research Unit FOR2783. The research leading to the presented results received additional funding from the European Regional Development Fund and the State of Thuringia via Thüringer Aufbaubank TAB (Contract No.~2019~FGI~0013).

\appendix

\section{Effects of large energy spreads on the source size estimation}
\label{appendix density contribution for different energies}

To illustrate the distinction in transverse momentum among electron beams with varying energy spreads, we conducted Particle-in-Cell (PIC) simulations using the SMILEI code~\cite{Derouillat.2018}. In these simulations, a particle beam with a mean energy of 75~MeV with various energy bandwidths, and an initial beam waist of $\sigma_x = 0.5$~\microm propagated through a laser grating produced by a laser of $\lambda_L=0.8$~\microm with a peak-to-peak distance of $\lambda_G=1.5$~\microm, a duration of 30~fs, and a strength of the grating of $a_0=0.6$. The momentum and divergence of the beam were recorded following the interaction.

\begin{figure}[ht]
    \centering\includegraphics[width=\columnwidth]{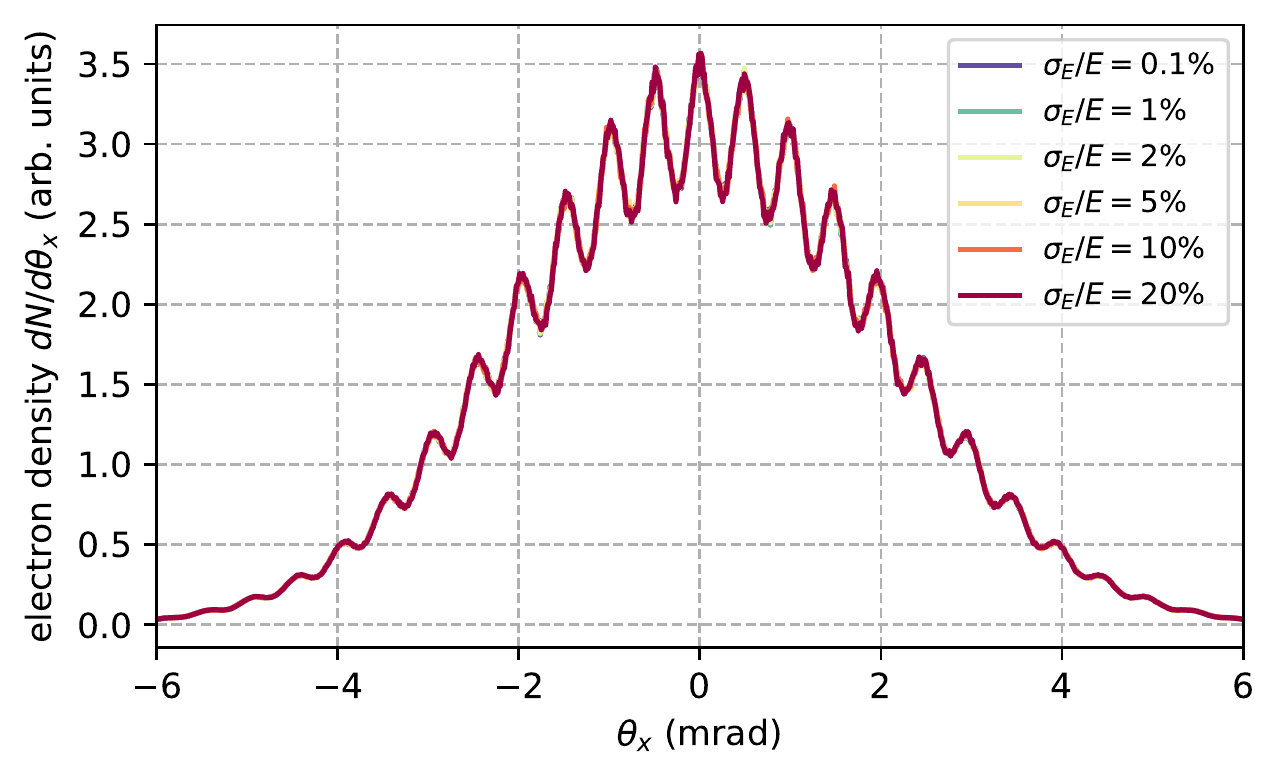}
    \caption{Simulated effects of different electron beam energy spreads $\sigma_E/E$ on the peak-to-valley modulations after interaction with the laser grating. No detectable difference in the modulated signal is observed, indicating that the variation in energy spread does not significantly impact the modulation.}
    \label{fig:energy spread effect}
\end{figure}

In Fig.~\ref{fig:energy spread effect}, the transverse momentum spaces of distinct electron bunches after the laser-particle interaction are illustrated. The simulation results indicate that, despite the varying energy spreads of the beam, the peak-to-valley modulations remain constant.

\begin{figure}
    \centering
    \includegraphics[width=\linewidth]{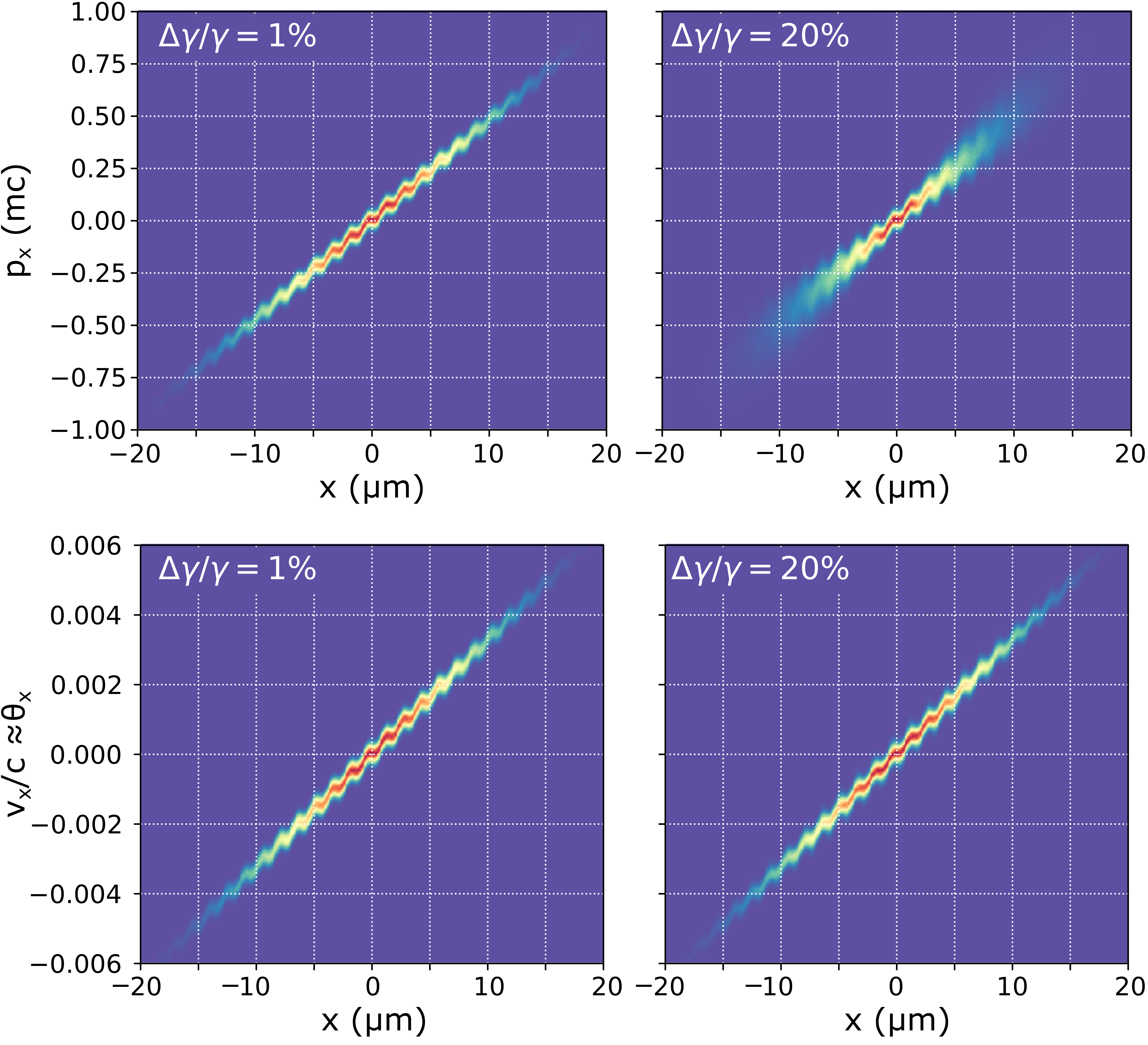}
    \caption{Simulation of the phase (upper panels) and trace (lower panels) spaces for beams of different energy spreads propagating though the same optical grating.}
    \label{fig:trace and phase space simulation}
\end{figure}

\rev{%
To understand why the modulations do not exhibit significant differences for various energy spreads, we can contrast the phase and trace spaces of the broadband beams after their interaction with the optical grating.

In the phase space of two different beams as depicted in Fig.~\ref{fig:trace and phase space simulation} (upper panels), we observe an increase in emittance for the beam with a large energy spread of 20\% (upper right panel) compared to the one with only a 1\% spread (upper left panel). Due to free drift, individual particles rotate with distinct velocities in phase space, leading to emittance growth in the broadband beam. Conversely, the beam with small energy spreads, including monoenergetic ones, maintains a relatively constant emittance. When both beams interact with the optical grating, the particles are subjected to a kick $\delta_{p_x}$ in transverse momentum, leading to the formation of modulations. However, for beams with large spreads, the modulations at the edge of the beam are smeared out since the emittance growth is more pronounced in this region.
}

\rev{%
However, the emittance growth in phase space due to free drift becomes insignificant in the trace space of the beam, as emphasized by Ref.~\cite{Antici.2012}. Consequently, the trace space ellipse exhibits negligible differences for both cases of different energy spreads. Hence, the kick, now described in terms of the particle velocity $\delta_{v_x}$, leads to an insignificant difference in the modulated signal observed on the detection screen during the experiment. This effect is consistent with the simulation results shown in the lower panels of Fig.~\ref{fig:trace and phase space simulation}.

Since the laser grating method relies on measuring the beam's divergence to reconstruct its trace space, the observed energy spread of the electron beam during the experiment does not result in any significant difference in the modulated signal compared to a monoenergetic beam. Hence, the theory and data analysis methods developed in this work are valid for our experimental conditions to infer the beam waist and rms emittance.
}%

\section{Interaction time between laser grating and electron beam}
\label{appendix interaction time}

\begin{figure}[b]
    \centering\includegraphics[width=0.9\columnwidth]{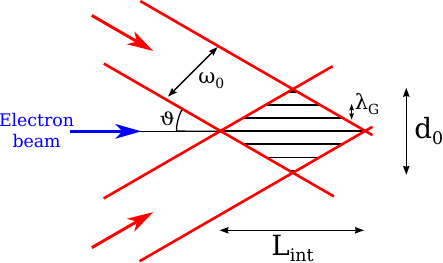}
    \caption{Sketch of the the interaction point between the laser grating and the electron beam. This schematic allows to estimate the interaction time between the optical grating and the electron beam.}
    \label{fig:crossing angle geometry}
\end{figure}

The interaction time between the laser grating and electron beam can be estimated by simply looking at the experimental geometry. Figure~\ref{fig:crossing angle geometry} shows a sketch of the collider laser beam with the interference pattern produced at focus. The half crossing angle of the laser (red) with respect to the electron beam (blue) is given by $\vartheta = 5.5^{\circ}$.

From the experimental sketch in Fig.~\ref{fig:crossing angle geometry}, the interaction time is given by $t_{\mathrm{int}} = L_{\mathrm{int}}/c$, where $c$ is the speed of light in vacuum. To determine $L_{\mathrm{int}}$, we simply apply trigonometry and find that $L_{\mathrm{int}} = d_0/ \tan \vartheta \approx 208~\si{\um}$, where $d_0 \approx 17~\si{\um}$ is the full width at half maximum of the grating pattern given previously in Figure~\ref{fig. pick-off laser beam}. Hence, the interaction time is  $t_{\mathrm{int}} \approx 582~\si{\fs}$.

\input{references.tex}

\end{document}